\documentclass{article}

\usepackage{mytemplate}

\usepackage{arxiv}
\usepackage[utf8]{inputenc} 
\usepackage[T1]{fontenc}    
\usepackage{hyperref}       
\usepackage{url}            
\usepackage{booktabs}       
\usepackage{amsfonts}       
\usepackage{nicefrac}       
\usepackage{microtype}      

\usepackage{graphicx}
\usepackage{natbib}
\usepackage{doi}

\usepackage{amssymb}
\usepackage{latexsym}

\newacronym{mis}{MIS}{medical image segmentation}

\newacronym{CT}{CT}{Computed Tomography}
\newacronym{MRI}{MRI}{Magnetic Resonance Imaging}
\newacronym{PET}{PET}{Positron-Emission Tomography}
\newacronym{CNN}{CNN}{convolutional neural network}
\newacronym{tn}{TN}{True Negative}
\newacronym{tp}{TP}{True Positive}
\newacronym{f1}{\text{F1}}{F1 measure}
\newacronym{fb}{\text{F\textbeta}}{F\textbeta\ measure}
\newacronym{fl}{FL}{Focal Loss}
\newacronym{fn}{FN}{False Negative}
\newacronym{fp}{FP}{False Positive}
\newacronym{fpr}{FPR}{False Positive Rate}
\newacronym{iou}{IoU}{Intersection over Union}

\definecolor{newcolor}{rgb}{.8,.349,.1}

\title{Boosting Medical Image Segmentation Performance with Adaptive Convolution Layer}
\author{
Seyed M.R. Modaresi\\
LIPN-UMR-CNRS 7030, \\
Sorbonne University Paris Nord,\\ Paris, France\\
\texttt{modaresi@lipn.univ-paris13.fr} \\
\And
Aomar Osmani \\
LIPN-UMR-CNRS 7030, \\
Sorbonne University Paris Nord,\\ Paris, France\\
\texttt{ao@lipn.univ-paris13.fr} \\
\And
Mohammadreza Razzazi\\
Computer Engineering Department, \\
Amirkabir University of Technology,\\ Tehran, Iran\\
\texttt{razzazi@aut.ac.ir}  \\
\AND
Abdelghani Chibani\\
Laboratory of Images, Signals and Intelligent Systems\\ Université Paris-Est Créteil,\\ Paris, France\\
\texttt{achibani@gmail.com}
}
\begin{document}

\maketitle




\title{Boosting Medical Image Segmentation Performance with Adaptive Convolution Layer}

\begin{abstract}
    Medical image segmentation plays a vital role in various clinical applications, enabling accurate delineation and analysis of anatomical structures or pathological regions. Traditional \glspl{CNN} have achieved remarkable success in this field. However, they often rely on fixed kernel sizes, which can limit their performance and adaptability in medical images where features exhibit diverse scales and configurations due to variability in equipment, target sizes, and expert interpretations.
    In this paper, we propose an adaptive layer placed ahead of leading deep-learning models such as UCTransNet, which dynamically adjusts the  kernel size based on the local context of the input image.
    By adaptively capturing and fusing features at multiple scales, our approach enhances the network's ability to handle diverse anatomical structures and subtle image details, even for recently performing architectures that internally implement intra-scale modules, such as UCTransnet.
    Extensive experiments are conducted on
    benchmark medical image datasets to evaluate the effectiveness of our proposal. It  consistently   outperforms traditional \glspl{CNN} with fixed kernel sizes with a similar number of parameters, achieving superior segmentation Accuracy, Dice, and IoU in popular datasets such as SegPC2021 and ISIC2018.  The model and data are published in the open-source repository, ensuring transparency and reproducibility of our promising results.
\end{abstract}

\keywords{
    Medical Image Segmentation\and Deep Learning\and
    Adaptive Convolution
}

\section{Introduction}
As a trending subject in the field of image processing and computer vision \citep{AsgariTaghanaki2021}, \Gls{mis} involves extracting the boundaries of desired targets, such as tumors, in medical images and determining  their class \citep{Luo2022}.
The accurate segmentation of medical images, such as \gls{CT}, \gls{PET}, and \gls{MRI}, plays a vital role in the diagnosis and treatment of various diseases and assists physicians in patient management, including staging, assessment, and prognosis of the treatment response \citep{Li2019,Liu2021,tian2021radiomics}.

Deep learning-based \gls{mis} has gained considerable traction in recent years \citep{Chen2023,Houssein2021,Devunooru2021,AsgariTaghanaki2021,Malhotra2022,Luo2022,Kaviani2022}. A myriad of models has been introduced in the literature for various \gls{mis} tasks and clinical outcomes, encompassing multi-organ detection, tissue mass detection, tumor or nodule segmentation and classification, cell counting, multiple diagnoses, prognosis, and the prediction of treatment outcomes for various chronic diseases like cancers or neurodegenerative diseases \citep{Kumar2022b,Devunooru2021,Roy2023,Simpson2019,Antonelli2021,Roth2016,Song2022a}.
In the context of cancer diagnosis, for instance, deep models such as \citep{melekoodappattu2022breast,Eali2022,Iqbal2022,Isensee2021,Singh2022b} have shown improved performance in segmenting tumors or nodules. Despite the significant progress, implementing deep learning in \gls{mis} continues to pose challenges since medical images often contain noise, artifacts, adhesions, and other distortions that can negatively affect the performance of deep learning models, particularly when discerning tumoral tissue boundaries and surroundings \citep{Kumar2022b,Malhotra2022,tian2021radiomics}.

\gls{CNN} models are widely used for analyzing and processing medical images. The fundamental block of \gls{CNN} is the convolution layer, which contains tiny adjustable weight grids (known as kernels) that convolved on the input image.
This work involves moving the kernel over the image, and at every stop, it performs a mathematical operation (dot product) using the weights of filters  and the pixel values of the image  to get a single output.
This process provides Local Receptive Fields for every pixel, assisting the model in concentrating on nearby characteristics. As we go deeper, the model starts recognizing more intricate and broader patterns, enhancing its understanding of the image \citep{Liu2021}.

\begin{figure}[t!]
    \centering
    \includegraphics[width=\columnwidth]{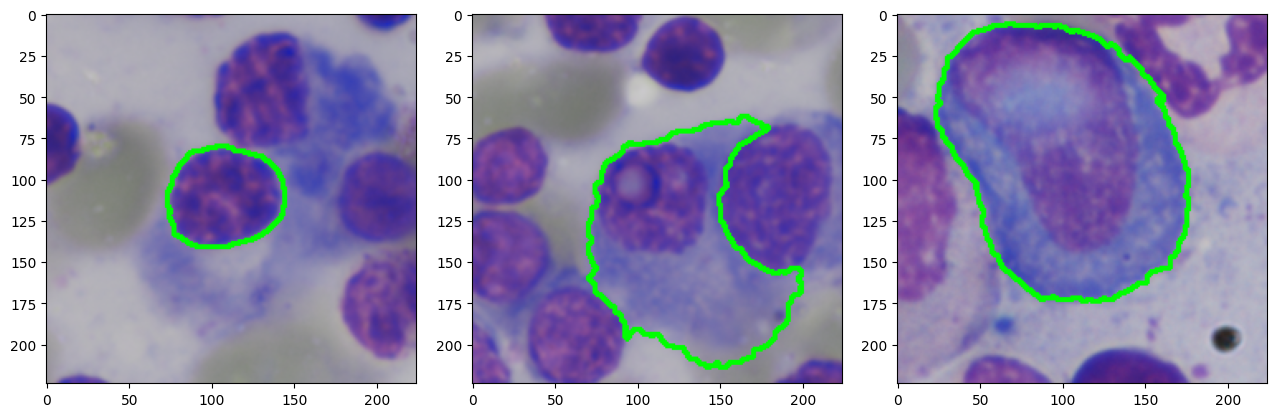}
    \caption{Illustration of diverse scales observed in medical images sourced from the SegPC-2021 dataset. The green contours highlight the regions containing cancerous lesions.}
    \label{fig:various scale medical images}
\end{figure}

Typically, kernels of a predetermined size are employed, often configured as a 3$\times$3 grid. This can be likened to a windowing technique where the kernel moves over the image in small segments, processing each segment to create a corresponding output in a new feature map. This process involves the kernel striding across the image, analyzing one small region at a time.
A number of research works highlight that the size of these receptive fields plays a crucial role in enhancing the efficiency of the models \citep{Ibtehaz2020,Milletari2016,Wang2021b,Qiu2018}. Consequently, sticking to a kernel of a fixed size might not work well for every image. For instance, larger input images or targets might require bigger receptive fields. As demonstrated in \cref{fig:various scale medical images}, having varying scales in different images suggests the usefulness of having diverse receptive fields to accommodate all of them \citep{Ibtehaz2020}.

To address these challenges, this paper focuses on developing an adaptive layer to be placed ahead of leading deep models, such as UCTransNet  architecture for \gls{mis}. The proposed model aims to overcome the limitations imposed by fixed kernel sizes in traditional \glspl{CNN} in the first layer, which may not adequately handle diverse anatomical structures, irregular shapes, and varying feature scales in medical images. It dynamically adjusts the kernel size based on the local context of the input image. This adaptive mechanism enables the network to capture relevant multiple scale features from the original input image before feeding to the inner network, leading to improved segmentation accuracy and robustness.

In the experiments, we conduct an extensive evaluation of this architecture using benchmark medical image datasets (ISIC2018 and SegPC2021 described in \cref{sec:datasets}). By comparing the results with traditional \glspl{CNN} that employ fixed kernel sizes, we demonstrate the superior performance and generalizability of our adaptive approach.
To ensure transparency, reproducibility, and consistency in implementation, our method and used datasets, including the state-of-the-art deep models,
are published in our open-source repository\footnote{\label{repository}\url{https://github.com/modaresimr/adaptive_mis}}.

The rest of the paper is organized as follows: Section 2 provides an overview of the related work in medical image segmentation and mutli-scale CNN architectures. Section 3 describes the methodology and architecture of the proposed architecture in detail. Section 4 presents the experimental setup, including the datasets, evaluation metrics, the results and analysis, followed by the conclusion and directions for future research in Section 5.

\section{Related Works}
The well-known U-Net model proposed by \citet{Ronneberger2015} gained significant  attention and is an influential architecture in the field of deep learning.
Similar to the Autoencoder models, the U-Net model contains the encoder (contracting)  and the decoder (expanding) paths.
The unique feature of U-Net is the incorporation of skip connections that enable the flow of information from the encoder to the decoder at different scales, facilitating the preservation of spatial details and improving the localization accuracy of the segmentation results \citep{Ronneberger2015,Azad2022}.
Numerous extensions of U-Net have been proposed to improve recognition quality in  medical tasks.
\citet{Azad2022} provide a comprehensive survey on U-Net and categorize the U-Net extensions into \textit{Skip Connection Enhancements}, \textit{Backbone Design Enhancements}, \textit{Bottleneck Enhancements}, \textit{Transformers}, \textit{Rich Representation Enhancements}, and \textit{Probabilistic Design}.
The MISSFormer model \citep{Huang2022b} redesigns the U-Net architecture by incorporating a position-free and hierarchical U-shaped transformer. It
utilizes the Enhanced multi-scale Transformer  module to bridge the gap between the encoder and decoder feature maps.  It has a slightly higher performance in the Synapse dataset \citep{Azad2022}.

The recent UCTransNet \citep{Wang2022l} proposes replacing simple skip connections in U-Net with a multi-scale channel-wise module to solve the semantic gaps for an accurate \gls{mis}. It also includes  an attention mechanism and transformer sub-module. The attention mechanism implicitly learns to suppress irrelevant regions while emphasizing  the regions of interest, while the transformer aids in capturing long-range dependencies and addresses the limitation of local receptive fields. The transformer sub-module tokenizes feature maps in  each stage within the appropriate patch sizes.

In addition to altering skip connections, transformers, and attention mechanisms,  alternative backbones are commonly used to improve U-Net performance.
ResNet \citep{Cicek2016} is also a common backbone for the U-Net architecture which addresses the issues of stacking many layers in deep neural networks that causes vanishing gradient problem.
The Google inception module, widely utilized for extracting features across multiple scales, was initially introduced in InceptionV1, where kernels of different sizes were concatenated in parallel \citep{Szegedy2015,Szegedy2016,Szegedy2017,Zhang2021d}. This architecture underwent further refinements in subsequent versions, with InceptionV2 replacing the 5$\times$5 convolution with two stacked 3$\times$3 convolutions and InceptionV4 breaking down square convolutional kernels into two vectors to reduce computational operations while increasing the receptive field \citep{Szegedy2016,Szegedy2017,AlShoura2023}. However, this led to a limitation where larger kernels could not be broken down, resulting in fewer selectable features.
MultiRes blocks, which employ a series of convolutional layers with residual connections, have been utilized to provide features at different scales, although with limited efficacy for small images and fuzzy objects \citep{Ibtehaz2020,Hossain2023,Lou2021}. To overcome these limitations, the dual-channel UNet (DC-UNet) was proposed to incorporate more different-scale features  at the cost of increased network parameters and GPU memory consumption \citep{Lou2021,Ansari2022}.
\citet{Gridach2021,Jiang2019,Lou2022,Yang2020a,Fu2023,Wang2021c,Zhan2023} consider fixed numbers of features for each multi-scale dilated atrous receptive field in parallel to not increase the computational complexity, which is another way of representing information at various scales. This strategy increases the receptive field of the layer without adding additional network parameters.  A convolutional filter in a CNN can be decomposed as a linear combination of pre-fixed bases particularly Fourier-Bessel  bases \citep{Qiu2018}.
\citet{Wang2021b} combine the idea of adaptive convolutional kernel and  combination of pre-fixed bases and replace all convolution filters with adaptive atoms which are shown slightly better in Image Classification tasks particularly when dealing with intra-image variance, while it is not yet applied for image segmentation tasks.
\begin{figure*}
    \centering
    \includegraphics[width=\textwidth]{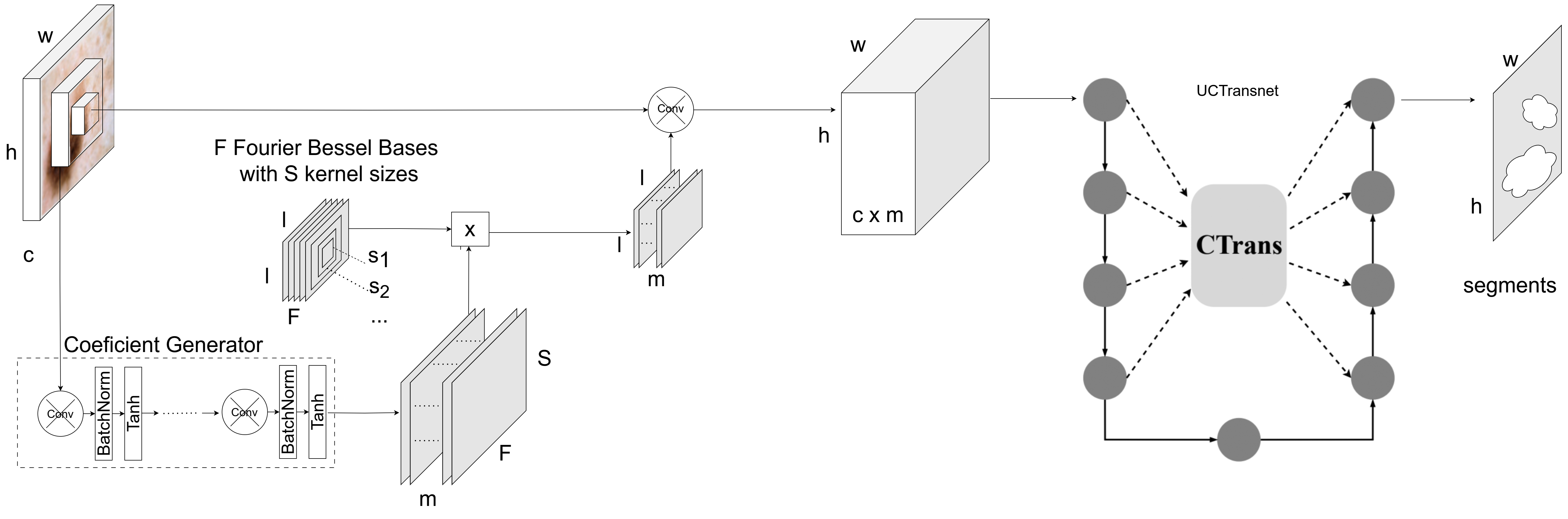}
    \caption{Adaptive Convolution Layer added to the leading UCTransNet architecture. The coefficient generator network generates the weights for Fourier-Bessel bases with different sizes for each  pixel and channel. It results a fixed kernel to be convolved for that pixel.}
    \label{fig:network}
\end{figure*}
In summary, recently many approaches are proposed to improve segmentation performance that are mainly concentrated on the improving skip connections, including attention mechanism and transformer and improving the backbone such as including dynamic convolution. The UCTransNet \citep{Wang2022l} performs better compare to other approaches such as TransUNet \citep{Chen2021d}, Residual U-Net \citep{Zhang2018b}, MultiResUNet \citep{Ibtehaz2020}, U-Net++ \citep{Zhou2018}, Att-UNet \citep{Oktay2018} and original U-Net \citep{Ronneberger2015} over several datasets such as ISIC 2018, and SegPC 2021 \citep{Azad2022}.

Although employing a dynamic receptive field offers theoretical advantages, it is still challenging to implement in practice. On one side, certain approaches necessitate extensive computational resources and memory to calculate the dynamic receptive field; on the other, they may not yield any enhancement in performance.
After a comprehensive study of various adaptive methods, the rest of this paper will present a novel dynamic receptive field layer that can be placed ahead of leading models. It improves the recognition performance while maintaining a similar number of parameters.

\section{Proposed Approach}

In this section, we present an adaptive convolution layer to be incorporated into leading models. Though it is not exclusive to UCTransNet, we showcase its integration at the top of this architecture, naming it AdaptUCTransNet.
The AdaptUCTransNet dynamically adjusts the kernel size based on the local context of the input image, enabling the network to capture relevant features at multiple scales from the first layer, leading to improved segmentation accuracy, robustness, and better consider diverse anatomical structures, irregular shapes, and varying feature scales in medical images.
Additionally, leveraging the benefits of the UCTransNet architecture, which integrates U-Net and transformer models, AdaptUCTransNet combines spatial information with self-attention mechanisms to extract meaningful features from medical images.

In \gls{mis}, the presence of diverse segments necessitates the adoption of a flexible approach. By adjusting the kernel size based on the specific context of each pixel, we can have a higher level of discernment when extracting features within the inner layers of deep networks.
By liberating the network from strict reliance on the hyperparameter associated with the dataset for determining the kernel size, we empower it to adapt and optimize its performance according to the unique characteristics of the input data.
Therefore, the proposed approach is to add the adaptive multi-size kernel convolution  layer to the best deep learning model for \gls{mis}.
We leverage the fact that convolution layers can be mathematically expressed as a linear combination of predetermined bases \citep{Qiu2018}. By employing a limited number of Fourier-Bessel bases, we substantially reduce the number of parameters. Notably, this reduction in parameters does not compromise the accuracy of image classification tasks \citep{Qiu2018,Wang2021}, while it has not yet been explored for \gls{mis}. Additionally, using  Fourier-Bessel bases improve the recognition of the structural information of the input image and  effectively mitigates the impact of high-frequency noise and addresses the computational complexities associated with employing multiple kernel sizes within the convolution layer.

Formally, the symbol $[p]$ is employed to denote the spatial coordinates on a feature map $Z$. These coordinates are applicable to a range of dimensional structures, extending from one-dimension to more complex, higher-dimensional forms.
The notation $\mathcal{N}^\delta_{Z[p]}$ is used to represent the receptive field surrounding the feature vector $Z[p]$, where the distance is within a $\delta$. The size of this receptive field can vary, ranging from being quite small to encompassing the entirety of the input feature.
Considering this receptive field, the $\mathcal{T}$ is used to represent  transformed inner receptive field. For instance, in the dilated receptive field, $\mathcal{T}$ selects a subset of input features at specified intervals. Conversely, in the case of a traditional convolution layer, $\mathcal{T}$ encompasses all pixels, incorporating them without any alterations.
Then, traditionally, this inner receptive field is convolved with a set of kernels, denoted as $K$. The convolution operation can be mathematically expressed as $Z'[p] = K * \mathcal{T}(\mathcal{N}^\delta_{Z[p]})$, where $Z'$ denotes the resultant feature map post-convolution, and $*$ operation sums the element-wise product.
i.e., this can be expressed as $K*N=\sum_q {(K\:\circ\:N)}_q$, where $\:\circ\:$ signifies Hadamard element-wise multiplication and the sum is taken over all elements of the product. These kernels are learned end-to-end by the network during the training process, typically via backpropagation and gradient descent. The inclusion of these kernels results in the addition of extra parameters to the network. As expected, using larger kernels further increases the total parameter count.
Therefore, it will be more challenging to dynamically select the most suitable kernel size for each spatial coordinate.

Based on \cite{Qiu2018}, a convolution kernel can be decomposed as a combination of Fourier-Bessel bases.
Therefore, instead of using a learnable kernel, we can learn the weights ($W$) for the pre-fixed Fourier-Bessel bases with different sizes ($FS$).
Therefore, $Z'[p] = FS\times W * \mathcal{T}(\mathcal{N}^\delta_{Z[p]})$.
For adaptively changing the receptive field, we use another inner network to learn these weights ($W$) based on the receptive field $\mathcal{N}^\delta_{Z[p]}$. This inner network, called the coefficient generator network, is trained end-to-end with the backpropagation and gradient descent.
In the process of selecting appropriate weights for kernels of varying sizes, we  stack several layers of smaller kernels to control the complexity, as suggested by \citet{Simonyan2015}. This approach ensures that the output comprehensively covers the entire receptive field. For example, by stacking a minimum of four layers of $3\times 3$ kernels, we can achieve $9\times 9$ receptive field.
This multi layer network convolved through the receptive field $\mathcal{N}^\delta_{Z[p]}$ with a smaller kernel size, and the output of this network is the local weights ($W({\mathcal{N}^\delta_{Z[p]}})$) for the pre-fixed Fourier-Bessel bases with different sizes ($FS$). This inner network remains fixed across all receptive fields. Therefore, this approach maintains the translation invariance characteristic of convolutional networks while taking into account the local context.

Building upon the specific attributes of medical images, we present our framework to enhance the accuracy of the best segmentation deep network.
The inclusion of our adaptive layer does lead to an increase in the number of parameters, but this increase constitutes only a minor fraction of the entire parameters of the network (\cref{sec:model complexity}).
A graphical representation of the network architecture is provided in \cref{fig:network}.
Given  set $F$ of Fourier-Bessel bases with $\abs{S}$ different sizes denotaed as $FS$,
for every pixel and channel another receptive field (local neighbors of the corresponding pixel which is greater than $S$) will be convolved in the coefficient generator network using smaller kernels. This network will be trained end to end in the training phase of the whole network. It produces $W$ with size $\abs{F} \times \abs{S}\times m$ weights, where $m$ is the number of intermediate channels. Matrix multiplication of Fourier-Bessel bases ($FS$) and these weights ($W$), generates a kernel for the given pixel and channel.
The adaptive layer convolves the input image with the explained kernel, resulting in $m$ intermediate features for each pixel and channel.
Then, these intermediate features will be fed  to the leading segmentation technique UCTransNet \citep{Wang2022l}, which replaces the simple skip connections in U-Net with a multi-scale channel-wise module. By using this combination, we can improve the performance of the model.

\section{Experiments}
In the experiments, we conduct an extensive evaluation of the adaptive layer added ahead of UCTransNet, AttUNet and well-known U-Net architecture using benchmark medical image datasets.
By comparing the results with traditional CNNs that employ fixed kernel sizes, we demonstrate the superior performance and generalizability of our adaptive approach.
Experiments are conducted  on various  public testbeds, including the Multiple Myeloma Plasma Cell Segmentation (SegPC) 2021 \citep{Gupta2023,gupta2021segpc} and the International Skin Imaging Collaboration (ISIC) 2018 datasets \citep{Codella2019}, which will be explained in details in the next sub section. Then, after explaining the details of implementation, we present the experimental results.

\subsection{Environment Setup}
In order to foster transparency and repeatability of our work, All the codes, datasets, and documentation are freely accessible on our GitHub repository: \url{https://github.com/modaresimr/adaptive_mis}.
All experiments are run on an NVIDIA DGX-1 machine featuring a Tesla V100\-32 GPU, Intel Xeon E5-2698v4 CPUs, and 512 GB of RAM. However, we use only a part of these resources.

\subsection{Datasets}
\label{sec:datasets}
Experiments are conducted  on various  public testbeds, including the Multiple Myeloma Plasma Cell Segmentation (SegPC) 2021 \citep{Gupta2023,gupta2021segpc} and the International Skin Imaging Collaboration (ISIC) 2018 datasets \citep{Codella2019}.
SegPC contains a collection of 775  microscopic 2D images from the bone marrow samples of Multiple Myeloma patients.
It significantly helped hematologists in making more accurate diagnoses and facilitates cancer screening.
The dataset from ISIC 2018 boasts a large collection of 2,594 RGB dermoscopy images. Robust segmentation of these images plays a crucial role in medical diagnosis and is challenging due to inconsistent lighting conditions, varying lesion sizes, texture disparities, and differences in color and positioning. Moreover, the presence of unrelated elements like air bubbles, hair strands, or ruler markers further add to the complexity \citep{Hasan2020,Codella2019,Gupta2023,gupta2021segpc,Azad2022}. Both datasets are illustrated in \cref{fig:comapre images,fig:isic comapre} with the segmentation result of our approach.

Similar to the work by \citet{Azad2022}, we allocated 70\% of images for training, 10\% for validation, and the remaining 20\% for testing, and our research focused on the segmentation of Cytoplasm components in SegPC 2021 and segmentation of cancer lesions in ISIS 2018 datasets.

\subsection{Hyperparameters and Implementation Details}
\label{sec:framework}
Our pipeline infers the segments from raw image data. All images underwent a sizing down operation to a standard size of $224 \times 224$ pixels. The pipeline is composed of an adaptive convolution layer with the kernel size of 3, 5, 7, and 9. We also select six Fourier Bessel bases similar to \cite{Wang2021b}.
For the coefficient generator network, we have used six intermediate features ($m$), which, is responsible for encoding the weights of the prefixed bases.
This network will be trained end to end during the global training process. We maintain early stopping with a patience of 20 epochs during the training. For better comparison, we make the other hyperparameters similar to the ones used in \citep{Azad2022}, such as the batch size of 16, epochs limit of 100,  Adam optimizer with a learning rate of 0.0001, and the average of cross-entropy loss and dice loss for the loss function.
The entire implementation, along with hyperparameters, is accessible and verifiable through our publicly available open-source repository.
\subsection{Model Complexity}
\label{sec:model complexity}
An essential factor in the assessment of models is the computational complexity. The number of trainable parameters of these components is listed in \cref{tab:parameters}.  Therefore, although  the training complexity is similar (the differences are less than 2\%), its performance is  better regarding to \cref{tab:comparision,tab:isic2018}.
\begin{table}[h!]
    \centering
    \caption{Number of million parameters in each model, including the adaptive variant. This indicates that although our method is effective in determining the ideal dynamic kernel size, it keeps the number of parameters almost the same as those of the original model.}
    \label{tab:parameters}
    \begin{tabular}{@{}lcclllllll@{}}
        \toprule
        Methods    & Normal & with Adaptive Layer \\ \midrule
        U-Net      & 19.487 & 19.850              \\
        Att-UNet   & 34.879 & 35.242              \\
        UCTransNet & 66.431 & 66.794              \\
        \bottomrule
    \end{tabular}
\end{table}
\subsection{Evaluation Metrics}
We used a series of performance metrics for a comprehensive analysis of our model's effectiveness. Precision and Recall served as the primary metrics. Precision assesses the model's ability to correctly predict positive instances, while Recall measures the completeness of these positive predictions.
Although Accuracy gives a general idea of the model's overall performance, it's crucial to interpret it alongside the other metrics due to potential data imbalance. We utilized the Intersection over Union (IoU) to measure the overlap between the predicted segmentation and the actual one.

As for the Dice Coefficient, it was used as an alternative to the F1 score due to its increased relevance in medical imaging. It places double emphasis on true positives which is the harmonic mean of precision and recall. It effectively gauges the spatial overlap of the predictions, which is particularly useful in biomedical image segmentation tasks.

\subsection{Experimental Results}

\begin{figure*}[t!]
    \centering
    \includegraphics[width=.9\textwidth]{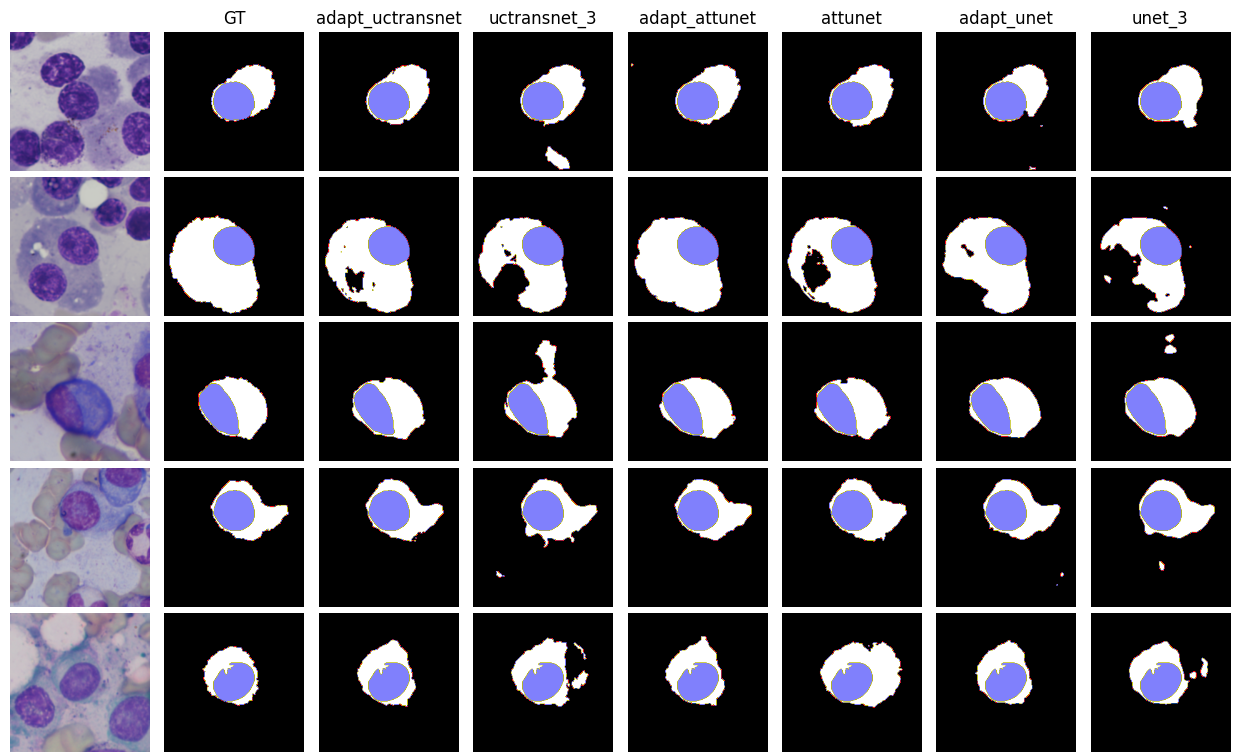}
    \caption{Visual comparisons of different methods for cytoplasm segmentation (depicted as the white region) on the SegPC 2021 dataset. The blue region denotes the Nucleus area of a cell. The initial column displays the input image, while the second column presents the ground truth. Following these, the subsequent columns feature the models along with their adaptive versions. As is evident, models incorporating the adaptive layer more accurately recognize the shape of the cytoplasm, and this improvement is particularly greater in larger segments. }
    \label{fig:comapre images}
\end{figure*}
\begin{figure*}[t!]
    \centering
    \includegraphics[width=.9\textwidth]{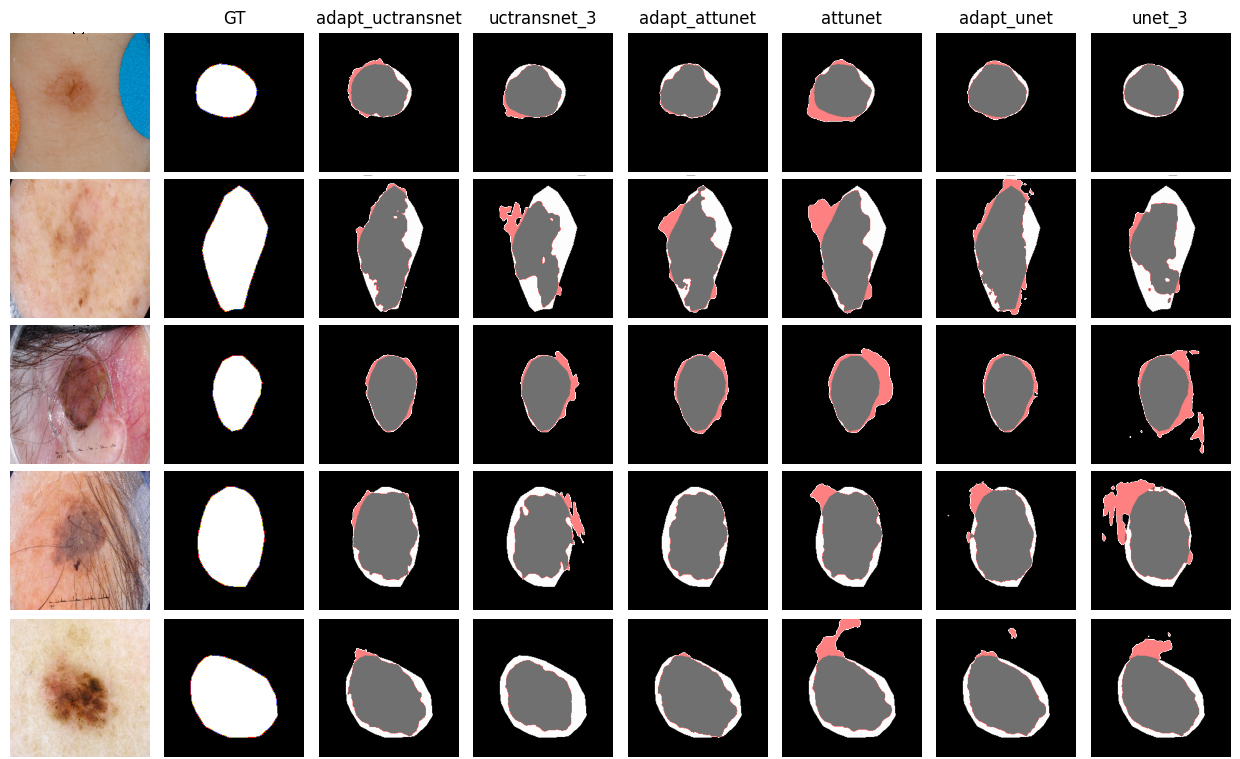}
    \caption{Segmentation output of various deep model in ISIC 2018 dataset. The white region represents the ground truth that remains undetected (\gls{fn}), while the gray region represents the detected ground truth (\gls{tp}), and red denotes the \gls{fp}. The columns orders are similar to \cref{fig:comapre images}. Once again, our model is more effective in identifying target regions, particularly noticeable in larger ones where traditional models with fixed kernels face difficulties in detecting intra-size features.}
    \label{fig:isic comapre}
\end{figure*}

\begin{figure}[t!]
    \centering
    \begin{minipage}[b]{0.47\textwidth}
        \centering
        \includegraphics[width=\textwidth]{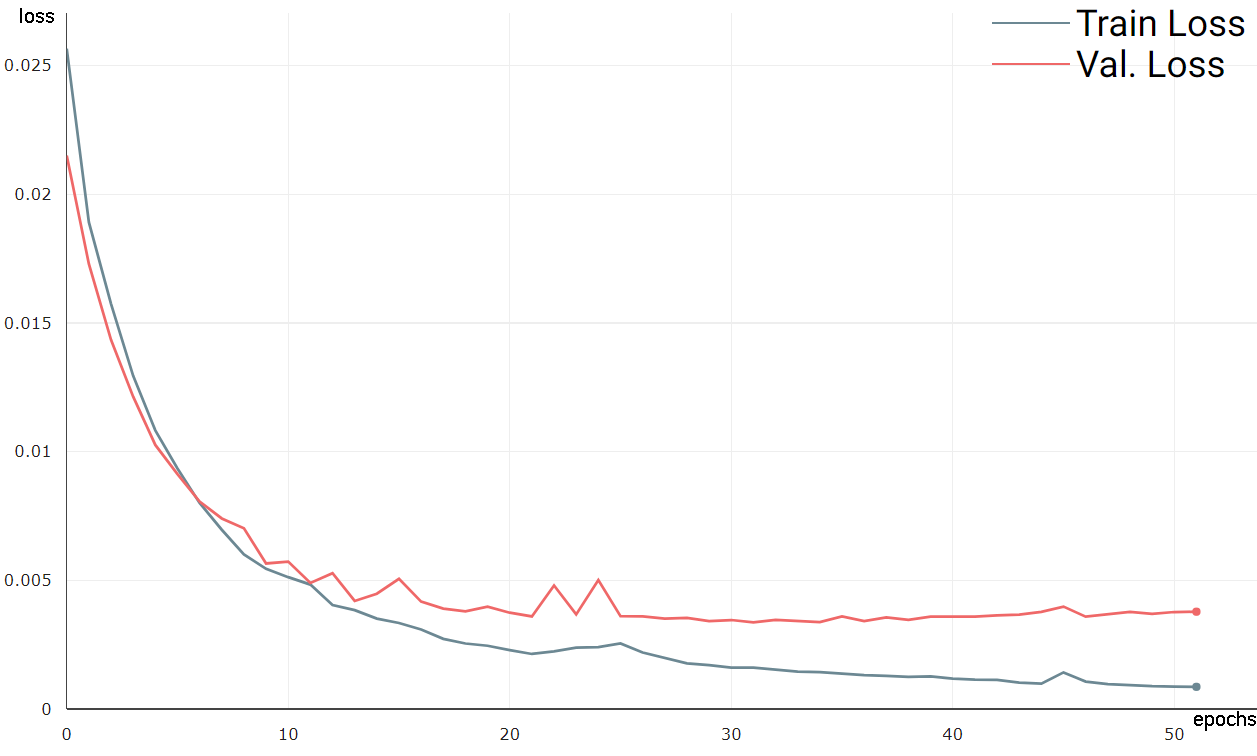}
    \end{minipage}
    \hfill
    \begin{minipage}[b]{0.47\textwidth}
        \centering
        \includegraphics[width=\textwidth]{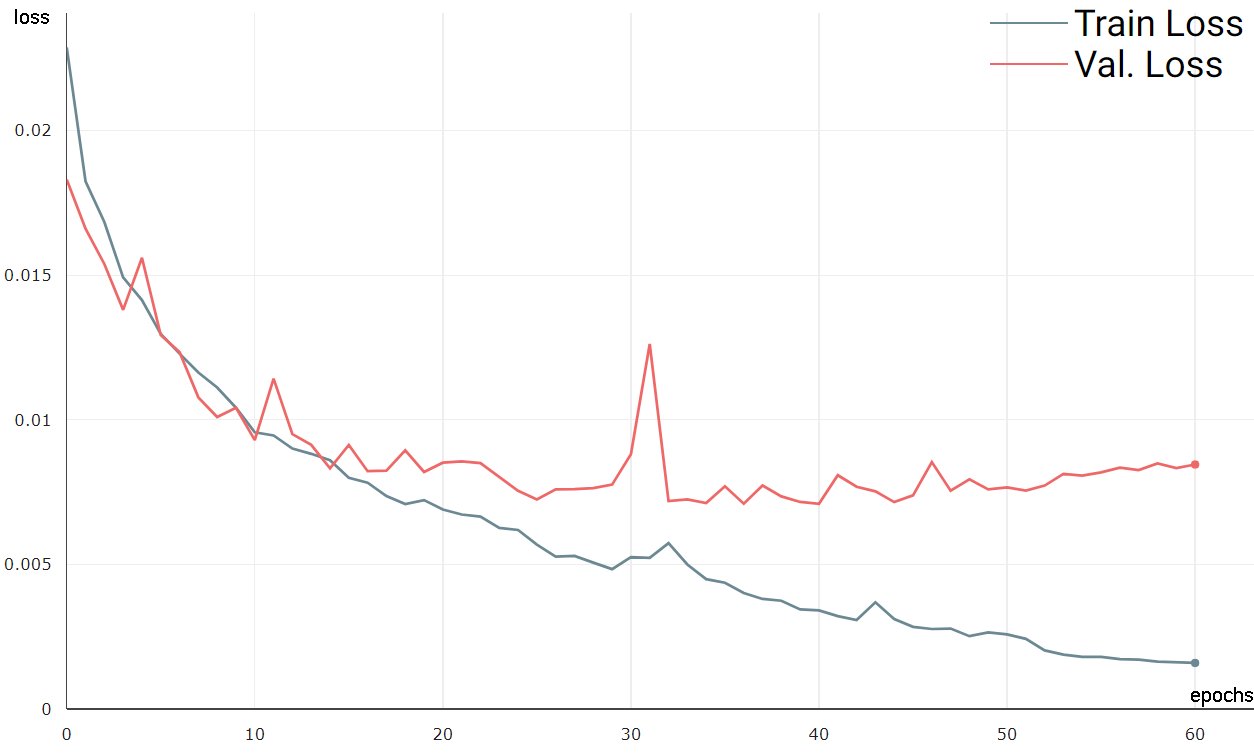}
    \end{minipage}
    \caption{The train loss and validation loss of SegPC 2021 dataset (left) and ISIC 2018 dataset (right). They indicate that the models are neither overfitting nor underfitting.}
    \label{fig:merged-captions}
\end{figure}

\subsubsection{Cell (SegPC 2021)}
We have showcased the visual segmentation results from the SegPC 2021 dataset in \cref{fig:comapre images}. Our adaptive multi-size-kernel representation effectively demonstrates its aptitude to generate  more accurate segmentation for cells of diverse scales and backgrounds.
A detailed comparison of results for the SegPC 2021 dataset is provided in \cref{tab:comparision}, further highlighting the effectiveness of our methodology. They demonstrate that incorporating our adaptive layer into the first layer of the model enhances the extraction of structural information, leading to a better virtual representation.
In \cref{fig:segpc2021-train__loss_validate__loss}, we have plotted the training and validation loss curves for the SegPC 2021 dataset.
These curves indicate the robust  performance of the model, as it is neither underfitting nor overfitting.

\begin{table}[t]
    \centering
    \caption{Comparison of results for the SegPC 2021 dataset. Each experiment is repeated five times. We have added our adaptive layer to two leading models (AttUNet, UCTransnet), and traditional UNet. As it is shown, even for architectures that internal include multi scale module, such as UCTransnet, adding our adaptive layer on top of that netwok improves not only the performance of these models but also their consistency (as shown by the standard deviation).
        For a more comprehensive comparison, other deep models, such as  Missformer, ResUNet, and MultiResUNet, are included at the second part of the table. Their models with our adaptive layer are accessible in our repository.}
    \label{tab:comparision}
    \small
    \begin{tabular}{llll}
        \toprule
        Model                      & Accuracy   & Dice       & IoU        \\
        \midrule
        \textbf{adapt\_UCTransnet} & 98.66±0.01 & 92.11±0.02 & 91.96±0.02 \\
        UCTransnet                 & 98.61±0.04 & 91.85±0.23 & 91.71±0.22 \\\textbf{adapt\_AttUNet} & 98.71±0.01 & 92.41±0.03 & 92.25±0.03 \\
        AttUNet                    & 98.65±0.02 & 92.10±0.08 & 91.95±0.08 \\

        \textbf{adapt\_UNet}       & 98.22±0.01 & 89.58±0.13 & 89.60±0.11 \\
        UNet                       & 98.07±0.05 & 88.69±0.35 & 88.80±0.30 \\ \hline
        Missformer                 & 98.35±0.04 & 90.38±0.16 & 90.32±0.15 \\
        ResUNet                    & 97.74±0.04 & 86.70±0.15 & 87.04±0.14 \\
        MultiResUNet               & 96.15±0.41 & 80.29±1.70 & 81.46±1.41 \\
        \bottomrule
    \end{tabular}
\end{table}

\subsubsection{Skin Cancer (ISIC 2018) dataset}

The segmentation result of our approach is illustrated in \cref{fig:isic comapre}.
Skin lesions typically manifest within the texture and seldom adhere to a definite shape or geometric pattern. This unpredictable behavior might explain why transformer-based networks may not yield substantial benefits for texture-related patterns \citep{Azad2022}.
Yet again, the adaptive  multi-size-kernel representation capability of our methodology demonstrates its proficiency. Compared to other approaches, it is remarkably effective at localizing abnormal regions, which is clearly illustrated in the segmentation results shown in \cref{fig:isic comapre}. This calls for a deeper exploration of the robustness and applicability of our approach. In \cref{fig:isic2018-train__loss_validate__loss_VS_epoch.jpeg}, we have plotted the training and validation loss curves for the ISIC 2018 dataset. The model is performing well, as indicated by the minimal gap (less than 0.003) between the training and validation losses and their stable behavior.

\subsection{Discussion}


The conducted experiments substantiate the efficacy of integrating an adaptive layer at the initial stage of deep networks, enhancing their resilience to diverse scales. This layer augments the network's capability to discern structural information across varying sizes, while maintaining a comparable parameter count.
The experiments demonstrate that the inputs including larger segments are better recognized by the proposed method. The noteworthy aspect of these experiments was the enhancement of all existing models through the integration of the adaptive layer, without necessitating any modifications to their structure. This improvement was observed even in models that inherently feature a multi-scale module (such as UCTransnet).
The accuracy of our experiment and number of parameters aligns with the recent survey by \citet{Azad2022}, further validating the credibility of our experimental results.

\begin{table}[t!]
    \centering
    \caption{Comparison of results for the ISIC 2018 dataset. It shows the effectiveness of our methodology. Similar to \cref{tab:comparision}, Each experiment is repeated five times and we have added our adaptive layer to two leading models (AttUNet, UCTransnet), and traditional UNet. This approach improves the performance of even the models with internal multi-scale module.
        For a more comprehensive comparison, other deep models, such as  Missformer, ResUNet, and MultiResUNet, are included at the second part of the table. Their models with our adaptive layer are accessible in our repository.}
    \label{tab:isic2018}
    \small
    \begin{tabular}{llll}
        \toprule
        Model                      & Accuracy   & Dice       & IoU        \\
        \midrule
        \textbf{adapt\_UCTransnet} & 95.64±0.13 & 89.31±0.18 & 87.68±0.16 \\
        UCTransnet                 & 95.54±0.07 & 89.04±0.27 & 87.40±0.25 \\
        \textbf{adapt\_AttUNet}    & 95.57±0.16 & 88.96±0.28 & 87.36±0.32 \\
        AttUNet                    & 95.44±0.15 & 88.66±0.25 & 87.04±0.29 \\
        \textbf{adapt\_UNet}       & 94.80±0.21 & 87.09±0.29 & 85.41±0.36 \\
        UNet                       & 94.43±0.25 & 86.18±0.39 & 84.49±0.46 \\ \hline
        Missformer                 & 95.25±0.18 & 88.38±0.42 & 86.69±0.43 \\
        ResUNet                    & 94.35±0.09 & 85.84±0.07 & 84.19±0.09 \\
        MultiResUNet               & 92.83±0.63 & 84.01±1.02 & 81.80±1.15 \\
        \bottomrule
    \end{tabular}
\end{table}

\section{Conclusion}
In this research, we delved into recent advancements in the field of computer vision, focusing on the dynamic modification of the receptive field, a strategy bearing resemblance to the previously described windowing approach. We introduced a novel adaptive layer designed for integration into \gls{mis}, aiming to enhance their overall performance.
We have shown the effectiveness of our adaptive layer approach  by including a dynamic layer on the top of the best segmentation deep network. This approach improves the recognition performance by dynamically changing the receptive field, resulting better identification of structural information and various size targets, and reducing high frequency noises while keeping the number of parameters nearly unchanged.
These promising results highlight the potential of our approach for broader applications in medical imaging.
Future research could explore adding the adaptive layer to the end of the network, potentially improving output processing, with initial signs suggesting further gains.
We also posit that the integration of an adaptive layer ahead of a lower-parameter inner network might achieve, or even surpass, the performance achieved with that higher-parameter inner network, that needs deeper investigation in future studies.


\section*{Funding Declaration}
This research did not receive any specific grant from public, commercial, or not-for-profit sectors. 
\section*{Declaration of competing interest}
The authors declare that they have no conflict of interest.

\section*{Data Availability Statement} The datasets, codes, and framework used in this study are accessible online through our repository at \url{https://github.com/modaresimr/adaptive_mis}.


\bibliographystyle{unsrtnat}
\bibliography{references}

\end{document}